# Understanding the Role of Cooperation in Emergency Plan Construction

**André Sabino**
Departamento de Informática,
Faculdade de Ciências e Tecnologia,
Universidade Nova de Lisboa,
Quinta da Torre, 2829-516 Caparica, Portugal
amgs@di.fct.unl.pt

**Armanda Rodrigues**
Departamento de Informática,
Faculdade de Ciências e Tecnologia,
Universidade Nova de Lisboa,
Quinta da Torre, 2829-516 Caparica, Portugal
arodrigues@di.fct.unl.pt

**ABSTRACT**

In this paper we describe a proposal for information organization for computer supported cooperative work, while working with spatial information. It is focused on emergency response plan construction, and the requirements extracted from that task.

At the centre of our proposal is the analysis of the structure of the cooperative workspace. We argue that the internal information representation should follow a spatial approach, tying the structure used to manage users with the structure used to manage information, suggesting the use of different spaces to represent the information.

The gain we expect from this approach is the improved capacity to extract information on how people are cooperating and their relationship with the information they are working with. The ideas are introduced while focusing on real life emergency planning activities, where we discuss the current shortcomings of the cooperation strategies in use and propose a solution.

**Keywords**

Emergency planning, computer supported cooperative work, geographic information systems.

**INTRODUCTION**

The work that leads to this paper is focused on the understanding of the way people cooperate over spatial information. As a case study, we looked at the planning activities that emergency management experts execute while preparing for potential emergency scenarios.

Given the different nature of the tasks, planning for emergencies has a different cooperation mechanics than dealing with a live event. Several authors recorded observations of how a control centre functions during a crisis, or drill (Mendonça, et al., 2009). For the present work, focussed on planning, we could not gather sufficient material to build a baseline of comparison on how people cooperate in such scenarios.

Therefore, we approached the Portuguese National Civil Protection Authority (ANPC), which is involved in the planning of major hazardous events in Portugal. The goal is to capture how people cooperate inside the ANPC, and how these officials relate with other agencies, when needed.

As a CSCW endeavour, we initiate by verifying if the problem is in fact a problem of cooperation (Briggs, 2006). Based on our previous work with the emergency management community, focused on the validation of emergency plans designed for dam break scenarios, we argue that such tasks and workspaces do require cooperation (Sabino, et al., 2008; Sabino and Rodrigues, 2009). From that experience, we concluded that, to design a useful emergency plan, several experts must be involved. This is due to the fact that the target area has to be analysed under different geographical dimensions and social concerns. These concerns are the responsibility of different agencies and organizations, which must have a say in the plan's construction.

There are also different ways one may be involved in plan construction. For instance, there is the direct input of elements on the plan's map, and the validation of such elements. Also, the granularity of the input varies from the specification of a single activity on the field to a whole scenario strategy definition (which involves the use of several lower level elements).

Furthermore, some plans legally require the input of several entities, as is the case with the Portuguese dam break emergency plans[1].

This work will document the planning procedures of experts. To do so, we will conduct a series of interviews with ANPC experts, designed to build a knowledge base from which we can draw conclusions about the cooperative process.

Being a cooperative effort, it is, then, interesting to understand how people organise themselves and how work is distributed among co-operators. How co-operators are kept aware of each other's input and processes, and how the work evolves through time, either by going through several iterations or by systematically adding higher-level constructions based on lower level ones.

**COMPUTER SUPPORTED COOPERATIVE WORK**

The computer science community has looked into cooperation with interest for a few decades now. From the initial assessment that computers can facilitate some people operated tasks, the interest on the automation of some cooperative endeavors rose.

The initial strategy of approaching the construction of computer systems for cooperation, with the same tools and mindset used in the design of single user applications failed. The reality is that cooperative work is in fact very different in nature when compared to single user work (Grudin, 1988).

Collaboration is generally understood as a process where a group of people cooperates to achieve a common goal, which co-exists with personal objectives. The nature of each particular contribution varies considerably, as does the responsibility held by each element, at a particular moment of the process. Both aspects involve dynamic behavior, enabling change in the role and interest scope of a contributor (Dourish and Bellotti, 1992).

Early examples on collaborative applications go from text editors (Leland, et al., 1992; Neuwirth, et al., 1988) and groupware applications (Greenberg and Marwood, 1994), to actual frameworks for the development of cooperative systems (Patterson, et al., 1990).

Although this work is focused on collaborative work over spatial domains, two main aspects should drive the design process of every collaboration support system: the concept of role and the notion of personal benefit (Grudin, 1988). Furthermore, the notion of awareness, i.e., the way a contributor perceives the actions of other collaborators, is also a key concept in the design of such systems (Dourish and Bellotti, 1992).

We could not find an existing project that could serve as comparison and benchmark. There are, however, some attempts to introduce cooperation into the spatial workspace. The project with most impact so far is a project by *ESRI*[2], named *Server GIS*, where information is available to more than one person. This is partially what we aim for, but lacks the support for cooperation at the user level. It only provides the means to make the information as available as possible, to as many people as desired.

There is also another project, *G2G ArcInfo*[3], where a peer-to-peer connection is enabled inside the ArcInfo platform. The users can chat and share a common map view. There is also the possibility to draw on the shared map. The approach is interesting because there are some new mechanisms for interaction alongside the traditional tools used to deal with spatial information. The major shortcoming is the fact that user management is unstructured, relying solely on the capacity of the users to organize outside the application, and thus remaining a provider of communication channels (other problems can also be identified, like the lack of history tracking, which can easily be addressed).

---

[1] Decreto-Lei no344/2007, Diário da República no198 Série I de 15 de Outubro de 2007 – (Portuguese law decree) from which one reads that every stakeholder on the dam construction and operation is required to provide input and cooperate with the necessary studies to assure safety to the structure and impact areas.

[2] *ESRI Server GIS* http://www.esri.com/technology-topics/server-gis/collaboration.html.

It is somewhat related with the ESRI *GeoPortal Server* open source project, aimed at spatial information service discovery http://www.esri.com/software/arcgis/geoportal/index.html

[3] *G2G ArcInfo* http://arcscripts.esri.com/details.asp?dbid=15578%00

**SPATIAL INFORMATION BASED COOPERATION**

Our main objective is to focus on the development of new ideas to support cooperation through the introduction of technology into the workspace. Specifically, we want to model applications that manage the spatial information that people use while planning a geographically relevant activity or otherwise use it to achieve some goal.

Each contribution to such a setup is either mapped on the hierarchy of cooperators or on the structure of spatial information, which is, in essence, geo-referenced. Using three sources of information, i.e., hierarchy and roles, information tagging and geo-referencing, we want to define a spatially measurable relationship between cooperators. The goal is to be able to evaluate how far a cooperator is to another, based on the calculated distance of the contributions.

We believe that, for the emergency response planning community, such a system would greatly improve the generation of cooperative links, because it enables the generation of suggestions for cooperation and hierarchical rearrangements. The ultimate objective is to produce a system for planning that uses a network of experts and plans, providing suggestions that lead to the construction of robust plans. It should also offer a workspace where the effort of finding the right expert for a specific issue is lowered by the suggestion mechanism, increasing the chances of actual cooperation.

We already produced a mock up of a possible workspace, where the most common channels for cooperation are present (direct communication and awareness), and with the suggestion mechanism integrated in the interface (see figure 1).

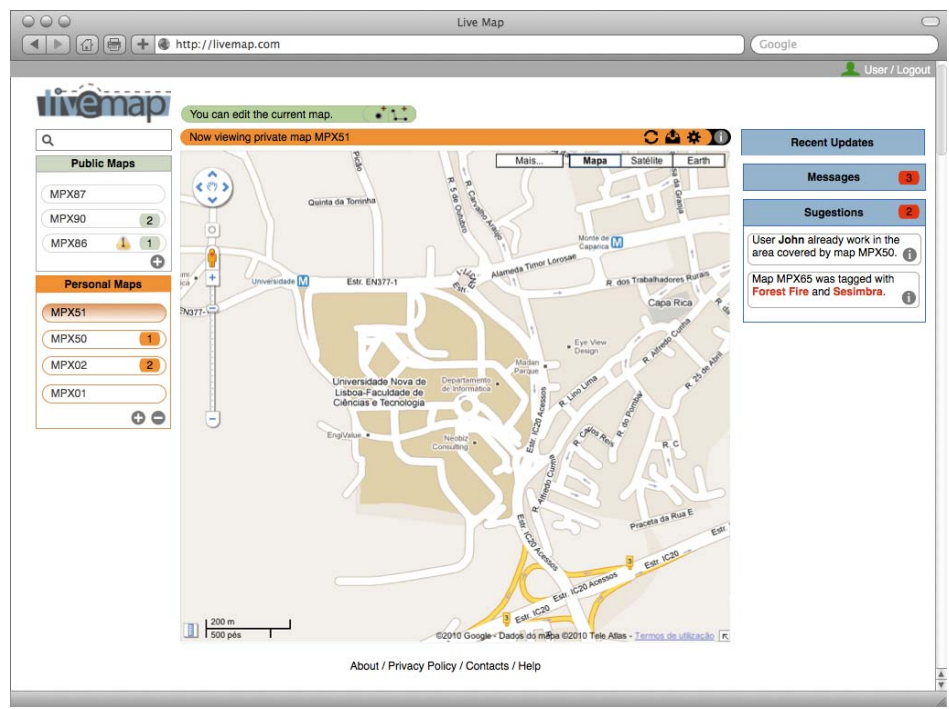

**Figure 1.** *livemap* **application prototype, illustrating the possibilities for a cooperative system with suggestions.**

At the core of the system is the suggestion mechanism. It is what we understand as lacking in the current plan construction process. Note that it is useful not only to create cooperative links, but also to create awareness about redundant work and missing elements of the plan (based on the analysis of tasks of previous plans).

Presently, the algorithm to manage the cooperative information is being defined. It is based on the three sources of information enumerated previously, and we expect to produce a mature version after the analysis of the knowledge base. The basis will enable us to balance the weights of the three components, and validate the output against examples we intend to build from the interviews. The questionnaire for the interviews is being designed with this goal in mind.

**COOPERATION KNOWLEDGE BASE**

The knowledge base we aim to build will capture several dimensions of the cooperative process. These are defined in table 1.

| Dimension | Description |
|---|---|
| Perception | To what degree the expert acknowledges that planning is a cooperative process? |
| Awareness | How is the expert aware of other planner's work? |
| Role Assignment | Is it clear which role the expert is meant to assume while planning? |
| Work Process | How is the workflow of plan construction? |
| Personal Benefit Assessment | What can the expert identify as potential gain from the use of technology to support cooperation? |

**Table 1. Dimensions of the cooperative process. Relates several relevant dimensions with the perceived reliance to the technological proposal.**

In table 1, the dimensions chosen to characterize the reality of the cooperative process, while planning for emergencies, are such that enable the clear perception of the process by the expert. A possible conclusion can be that the process is, currently, not at all cooperative. An extreme conclusion is that the perception of the experts is that it should not be.

For the case that it should be a cooperative effort, and assuming the first hypothesis (that it currently is not), and based on the whole characterization of the process, we propose an approach to cooperation that relates all the dimensions. Assuming the perception of experts clearly defines it as cooperative, we will focus on the deficiencies of the process, which in the past kept the plans from being constructed, at least to a state that could be subject to validation (Sabino and Rodrigues, 2009).

For the case that the input of the expert suggests that it should not be a cooperative effort, we will try to understand if such is caused by the lack of means to support that cooperative process. At the moment we can only support the idea of cooperation based on informal conversations with the ANPC and interactions with partners of previous work (Sabino, et al., 2008; Nóbrega, et al., 2008), hence the need to formally construct the knowledge base.

**INTERVIEWS**

The interviews will take place in early 2011. These are designed to meet the needs of our knowledge base.

The main guideline for conducting these interviews is to use the past experience of the experts to answer the questions referring to the various dimensions. Particularly, how past planning activities were conducted, either collaboratively or not, and what was the outcome.

**Current Record of Plan Construction**

The currently recorded information on plan construction is, to our best knowledge, non-existent. Most projects rely on partial observations and informal conversations, and mostly capture the personal opinion of a few experts. There are, in contrast, a collection of publications describing how control centers operate during an emergency and how some of the actors behave on the field (Mendonça, et al., 2009).

We can work with our current knowledge on how emergency planning is performed. In fact, the design of the mock up of the application was based on this. The assumption was that there is no single stakeholder that can design the whole plan; there is no current validation process for plan construction; It is not certain what could best fit a particular aspect of the plan; experts and technicians are interested in cooperation, but are not sure on how to actually support it; there is an interest in the record and documentation of plans.

The knowledge base will enable us to validate these assertions, and make design decisions accordingly.

**CONCLUSION**

Cooperation is present every time some task requires the input of several people. What we believe is that some cooperative efforts require the right tools to actually occur, namely geographically mapped activities. There is a threat to this statement: if some process does not happen in a work environment, should the sole introduction of computation enable it to occur?

To deal with that threat, we analysed the current processes of emergency plan construction and concluded that the outcome is not satisfactory. Looking at the task description and work environment, we believe that the task should be a cooperative effort. Currently, we must conclude that it is not, or, at least, it is not supported in a way that enables it to produce the desired outcome.

Based on these observations, we designed an application to support cooperative plan construction. Right now, we are in the position to record the opinion of emergency management experts, and compare it to our assumptions, tailoring the cooperation model used to outline the application.

With the construction of a cooperation knowledge base we want to validate our assumptions on how people cooperate in an emergency planning context, and use this record as a basis for future work.

**ACKNOWLEDGMENTS**

This work is partially funded by the Fundação para a Ciência e Tecnologia, doctoral grant SFRH / BD / 47403 / 2008.

We would like to thank the help and availability of the Portuguese National Civil Protection Authority (ANPC), particularly Patricia Pires, for the enabling contacts with experts and insight on emergency plan construction process.